\documentclass[preprint]{emulateapj}
\usepackage{epsf}
\usepackage{apjfonts}
\usepackage{xspace}
\usepackage{amsmath}
\begin{document}
\submitted{The Astrophysical Journal, to be submitted}
\slugcomment{{\em The Astrophysical Journal, to be submitted}} 
\shortauthors{LAU, KRAVTSOV, \& NAGAI}

\title{Residual Gas Motions in the Intracluster Medium\\ and Bias in Hydrostatic measurements of mass profiles of clusters}

\author{Erwin T. Lau\altaffilmark{1}, Andrey V.
  Kravtsov\altaffilmark{1,2}, Daisuke Nagai\altaffilmark{3,4}}



\altaffiltext{1}{Department of Astronomy \& Astrophysics, 5640 South Ellis Ave., The
  University of Chicago, Chicago, IL 60637 ({\tt ethlau@oddjob.uchicago.edu})} 
\altaffiltext{2}{Kavli Institute for Cosmological Physics and 
  Enrico Fermi Institute, 5640 South Ellis Ave., The University of
  Chicago, Chicago, IL 60637}
\altaffiltext{3}{Department of Physics, Yale University, New Haven, CT 06520}
\altaffiltext{4}{Yale Center for Astronomy \& Astrophysics, Yale University, New Haven, CT 06520}

\begin{abstract}

We present analysis of bulk and random gas motions in the intracluster medium
using high-resolution Eulerian cosmological simulations of sixteen
simulated clusters, including both very relaxed and unrelaxed systems
and spanning a virial mass range of $5\times 10^{13}-2\times
10^{15}\,h^{-1}\,{\rm M_{\odot}}$. We investigate effects of the
residual subsonic gas motions on the hydrostatic estimates of mass
profiles and concentrations of galaxy clusters.  In agreement with
previous studies we find that the gas motions contribute up to
$\approx 5\%-15\%$ of the total pressure support in relaxed clusters
with contribution increasing with cluster-centric radius. The
fractional pressure support is higher in unrelaxed systems.  This
contribution would not be accounted for in hydrostatic estimates of
the total mass profile and would lead to systematic underestimate of
mass. We demonstrate that total mass can be recovered accurately if
pressure due to gas motions measured in simulations is explicitly
taken into account in the equation of hydrostatic equilibrium. Given
that the underestimate of mass is increasing at larger radii, where
gas is less relaxed and contribution of gas motions to pressure is
larger, the total density profile derived from hydrostatic analysis is
more concentrated than the true profile. This may at least partially
explain some high values of concentrations of clusters estimated from
hydrostatic analysis of X-ray data.

\end{abstract}

\section{Introduction}

Clusters of galaxies are powerful cosmological probes and have the
potential to constrain properties of dark energy and its evolution
\cite[e.g.,][]{henry_arnaud91,holder_etal01,majumdar_mohr03,hu03,albrecht_etal06,vikhlinin_etal09b}.
Most of the cosmological applications using clusters rely on the
estimates of their total virial mass --- a quantity which is difficult
to measure accurately in observations. Traditionally, the masses are
estimated using observable X-ray properties of the intracluster plasma
using spectroscopic X-ray temperature or luminosity \cite[e.g.,][for a
review]{rosati_etal02} or velocity dispersion of cluster galaxies.
More recently, mass estimates via strong lensing \citep{smith_etal05},
weak lensing \citep[e.g.,][]{dahle06,mahdavi_etal08,zhang_etal08},
combined strong and weak lensing
\citep{brada_etal06,limousin_etal07,oguri_etal09}, and
Sunyaev-Zel'dovich effect \cite[e.g.,][for a recent
review]{carlstrom_etal02} have become available.

One of the most widely used methods for measuring cluster masses
utilizes gas density and temperature profiles of the intracluster
medium (ICM) derived from X-ray observations to solve the equation of
hydrostatic equilibrium (HSE) for the total mass profile under
assumptions of spherical symmetry and equilibrium
\citep[e.g.,][]{sarazin86,evrard_etal96}. With the advent of
high-resolution observations from the {\em Chandra} and {\em
XMM}-Newton satellites, which allowed measurements of the density and
temperature profiles to large radii, it became possible to use this
method without simplifying assumptions of isothermality and
$\beta$-model for gas distribution \citep[e.g.,][]{vik06}.

Recent tests using cosmological simulations of clusters have shown
that with resolved profiles and with reliable subtraction of the X-ray
background, the HSE method can derive masses of relaxed clusters with
an accuracy of better than $\approx 10\%-20\%$
\citep{rasia_etal06,nag07a}.  At the same time, simulations uniformly
show the presence of ubiquitous subsonic flows of gas even in very
relaxed clusters
\citep{evrard90,norman_bryan99,nag03,rasia_etal04,kay_etal04,faltenbacher_etal05,dolag_etal05,rasia_etal06,
nag07a,piffaretti_valdarnini08,jeltema_etal08,iapichino_etl08,ameglio_etal09}.

Such gas motions are thought to be driven by continuing accretion of
gas onto clusters along filaments, mergers and supersonic motions of
galaxies through the ICM.  Shocks can leave behind wakes
with sizes comparable to the length scale of the cluster.  Energy of the large-scale eddies in the flow can
cascade down to smaller scales. Gas motions on smaller scales can also be driven
directly by motions of groups and galaxies \citep{kim_07} and by jets and bubbles
from the active galactic nuclei \citep[AGN,][]{churazov_etal02},
although the latter may not be as important energetically as mergers
in contributing to gas motions and their contribution may
be confined to the inner regions of clusters.

Given that only thermal pressure is taken into account in the HSE
analysis, presence of random gas motions (or any other non-thermal
pressure component) can contribute to the pressure support in clusters
and bias HSE measurements of the total mass profiles \citep{evrard90}.
Analyses of simulated clusters show that up to $\approx 10\%-20\%$ of
pressure support comes from subsonic random gas motions of gas
\citep{rasia_etal04,faltenbacher_etal05,rasia_etal06}.  A recent
comparison of mass from weak lensing and X-ray HSE mass measurements by
\citet{mahdavi_etal08} shows that the HSE mass is biased low by 20\%
compared to lensing measurement suggesting that a non-hydrostatic
component in the gas pressure gradient \citep[see, however,][]{zhang_etal08}.

Random gas motions can alter the structure of the ICM through the
redistribution of energy from the decay of large-scale gas flows with a
time scale on the order of a few turnover time of major
mergers. In particular, it can also be responsible for dispersing
metals from the ICM core, where the abundance profile is broader than
the central galaxy brightness profile \citep{rebusco_etal05} and more
generally in mixing gas of different metallicity and entropy
\citep{wadsley_etal08,mitchell_etal09}. Incomplete thermalization of
gas motions results in lower ICM temperatures for clusters of a given
mass, which can contribute to the bias and scatter in the cluster
mass-temperature relation. Viscous dissipation of gas motions
can lead to secular heating of the ICM. Random gas motions can  
also maintain and amplify cluster magnetic fields via dynamo processes
\citep{roettiger_etal99,subramanian_etal06} and contribute to the
acceleration of cosmic rays in the ICM \citep[e.g.,
][]{brunetti_lazarian07}. It is therefore important to understand
gas flows in the ICM in order to understand both thermal and
non-thermal processes in the ICM.

Direct measurements of gas velocities via X-ray
spectroscopy \citep{inogamov_sunyaev03} are challenging with the current
X-ray instruments, but there
are indirect indications that residual gas motions
are indeed present in the ICM of observed clusters.
In particular, \citet{schuecker_etal04} use the Fourier analysis of 
fluctuations in the projected gas pressure map of the 
Coma Cluster measured with {\em XMM-Newton} and find a Kolmogorov-like
spectrum. Given that the properties of gas 
motions and the Reynolds number of the ICM in observed clusters are poorly known, cosmological 
simulations of cluster formation remain the best tool 
for studying properties of gas motions and evaluating 
their possible effect on observable properties of clusters.

In this paper we focus on quantifying the contribution of random gas
motions to pressure support in clusters and the corresponding bias in the HSE
estimates of total mass profile using a suite of high-resolution
cosmological simulations of cluster formation.  Our simulations
properly treat both collisionless dynamics of dark matter and stars
and gasdynamics in a self-consistent cosmological setting and capture
a variety of physical phenomena from the nonlinear collapse and
merging of dark matter to shock-heating and radiative cooling of gas,
star formation, chemical enrichment of the ICM by supernova and energy
feedback. These simulations should therefore faithfully capture the
dynamical processes driving ICM motions and can give us useful
insights into its expected effects.  Although a number of recent
studies have examined random gas motions and their effect on mass
estimate \citep{rasia_etal04,kay_etal04,dolag_etal05,rasia_etal06}, 
most of the studies have used simulations with SPH gasdynamics.  
The magnitude and effects of gas motions in such simulations depends 
on the specific treatment of artificial viscosity \citep{dolag_etal05}.  
Our study, which employs simulations with Eulerian gasdynamics with 
very low numerical viscosity, is therefore useful in evaluating possible 
differences between numerical techniques and systematic theoretical uncertainties.

In agreement with previous studies, we find that gas motions
contribute up to $\gtrsim 5-15\%$ of the pressure support and leads to
bias in the HSE mass measurement of a similar magnitude. This pressure
contribution and mass bias increase with cluster-centric radius and
for unrelaxed systems. The increasing underestimate of mass at larger
radii results in derived total density profile that is more concentrated than
the true profile in the simulation. This leads to an overestimate of
concentrations by $20\%$ when the derived profiles are fit with the
NFW profile \citep{nfw96}. This can at least partially explain high concentrations found
in some recent observational studies
\citep{maughan_etal07,buote_etal07}.  We also demonstrate that total
mass and density profiles can be recovered accurately if random gas
pressure is explicitly taken into account.

The paper is organized as follows. In \S~\ref{sec:simulations} we
present and describe our simulation sample of clusters. In Section 3
we analyze the velocity structure of gas and dark matter in the
simulated clusters, and give the results of the relative fractions of
random gas pressure and its gradient, and the expected bias in
hydrostatic cluster mass estimation. In Section 4 we summarize and
discuss our findings.

\section{The Simulations}
\label{sec:simulations}

In this study, we analyze high-resolution cosmological simulations of
16 cluster-sized systems in the flat {$\Lambda$}CDM model:
$\Omega_{\rm m}=1-\Omega_{\Lambda}=0.3$, $\Omega_{\rm b}=0.04286$,
$h=0.7$ and $\sigma_8=0.9$, where the Hubble constant is defined as
$100h{\ \rm km\ s^{-1}\ Mpc^{-1}}$, and an $\sigma_8$ is the power
spectrum normalization on an $8h^{-1}$~Mpc scale.  The simulations
were done with the Adaptive Refinement Tree (ART)
$N$-body$+$gasdynamics code \citep{kra99,kra02}, a Eulerian code that
uses adaptive refinement in space and time, and (non-adaptive)
refinement in mass \citep{klypin_etal01} to reach the high dynamic
range required to resolve cores of halos formed in self-consistent
cosmological simulations. The simulations presented here are discussed
in detail in \citet{nag07a} and \citet{nag07b} and we refer the reader
to these papers for more details. Here we summarize the main
parameters of the simulations.

High-resolution simulations were run using a uniform 128$^3$ grid and 8 levels
of mesh refinement in the computational boxes of $120\,h^{-1}$~Mpc for
CL101--107 and $80\,h^{-1}$~Mpc for CL3--24 (see Table~\ref{tab:sim}). These
simulations achieve a dynamic range of $32768$ and a formal peak resolution of
$\approx 3.66\,h^{-1}$~kpc and $2.44\,h^{-1}$~kpc, corresponding to the actual
resolution of $\approx 7\,h^{-1}$~kpc and $5\,h^{-1}$~kpc for the 120 and
$80\,h^{-1}$~Mpc boxes, respectively.  Only the region of $\sim
3-10\,h^{-1}$~Mpc around the cluster was adaptively refined, the rest of the
volume was followed on the uniform $128^3$ grid.  The mass resolution, $m_{\rm
  part}$, corresponds to the effective $512^3$ particles in the entire box, or
the Nyquist wavelength of $\lambda_{\rm Ny}=0.469\,h^{-1}$ and $0.312\,h^{-1}$
comoving Mpc for CL101--107 and CL3--24, respectively, or $0.018\,h^{-1}$ and
$0.006\,h^{-1}$~Mpc in the physical units at the initial redshift of the
simulations. The dark matter (DM) particle mass in the region around the
cluster was $9.1\times 10^{8}\,h^{-1}\, {\rm M_{\odot}}$ for CL101--107 and
$2.7\times 10^{8}\,h^{-1}\,{\rm M_{\odot}}$ for CL3--24, while other regions were
simulated with lower mass resolution.

To gauge the possible effects of resolution on our results, we have
used re-simulations of one of the clusters (CL6) with varying maximum refinement
levels from 6 to 9 (corresponding to difference in peak resolution of a factor of 8).

\begin{table}
\begin{center}
\caption{Properties of the simulated clusters at $z=0$.}\label{tab:sim}
\begin{tabular}{l c c c c  }
\hline
\hline
Cluster ID\hspace*{5mm} & 
{$M_{500c}$} & 
{$r_{500c}$} & 
{$v_{500c}$} & 
relaxed (1) \\
& {(10$^{14}$ $h^{-1} {\rm M_{\odot}}$)}
& {($h^{-1}$ Mpc)} 
& {(km s$^{-1}$)}
& unrelaxed (0) \\
\hline
CL101 \dotfill & 9.02 & 1.16 & 1830 & 0 \\
CL102 \dotfill & 5.45 & 0.98 & 1547 & 0 \\
CL103 \dotfill & 5.70 & 0.99 & 1571 & 0 \\
CL104 \dotfill & 5.40 & 0.98 & 1543 & 1 \\
CL105 \dotfill & 4.86 & 0.94 & 1489 & 0 \\
CL106 \dotfill & 3.47 & 0.84 & 1332 & 0 \\
CL107 \dotfill & 2.57 & 0.76 & 1205 & 0 \\
CL3 \dotfill & 2.09 & 0.71 & 1125 & 1 \\
CL5 \dotfill & 1.31 & 0.61 & 964 & 1 \\
CL6 \dotfill & 1.68 & 0.66 & 1046 & 0 \\
CL7 \dotfill & 1.42 & 0.63 & 989 & 1 \\
CL9  \dotfill & 0.83 & 0.52 & 826 & 0 \\
CL10 \dotfill & 0.67 & 0.49 & 770 & 1 \\
CL11 \dotfill & 0.90 & 0.54 & 847 & 0 \\
CL14 \dotfill & 0.77 & 0.51 & 806 & 1 \\
CL24 \dotfill & 0.35 & 0.39 & 619 & 0 \\
\hline
\end{tabular}
\end{center}
\end{table}

The $N-$body$+$gasdynamics cluster simulations used in this analysis
include collisionless dynamics of DM and stars, gasdynamics and
several physical processes critical to various aspects of galaxy
formation: star formation, metal enrichment and thermal feedback due
to supernovae Type II and Type Ia, self-consistent advection of
metals, metallicity dependent radiative cooling and UV heating due to
cosmological ionizing background \citep{haardt_madau96}.  The cooling
and heating rates take into account Compton heating and cooling of
plasma, UV heating, and atomic and molecular cooling, and are
tabulated for the temperature range $10^2<T<10^9$~K and a grid of
metallicities, and UV intensities using the {\tt Cloudy} code
\citep[ver.  96b4;][]{ferland_etal98}.  The Cloudy cooling and heating
rates take into account metallicity of the gas, which is calculated
self-consistently in the simulation, so that the local cooling rates
depend on the local metallicity of the gas. Star formation in these
simulations was done using the observationally-motivated recipe
\citep[e.g.,][]{kennicutt98}: $\dot{\rho}_{\ast}=\rho_{\rm
gas}^{1.5}/t_{\ast}$, with $t_{\ast}=4\times 10^9$~yrs. The code also
accounts for the stellar feedback on the surrounding gas, including
injection of energy and heavy elements (metals) via stellar winds,
supernovae, and secular mass loss.  

These simulations therefore follow the formation of galaxy clusters
starting from the well-defined cosmological initial conditions and
capture the dynamics and properties of the ICM in a realistic
cosmological context.  However, some potentially relevant physical
processes, such as AGN bubbles, magnetic field, and cosmic rays, are
not included.  Therefore, the simulated ICM properties are probably
not fully realistic in the innermost cluster regions, where these
processes are likely important. Moreover, the gas in the simulations
is treated as an ideal inviscid fluid with a small amount of numerical
Lapidus viscosity,\footnote{This term is
much smaller than the usual artificial viscosity term employed in the
SPH simulations and is not important dynamically.} and it remains unclear to what extent the level of motions in the ICM
 found in the simulations and discussed below applies to
real clusters.  Despite these limitations, the current simulations
reproduce the observed ICM profiles outside cluster cores
\citep{nag07b} and are therefore sufficiently realistic for a purpose
of the current study.

Our simulated sample includes 16 clusters at $z=0$ and their most massive
progenitors at $z=0.6$. The properties of simulated clusters at $z=0$ are
given in Table~\ref{tab:sim}. The total cluster masses are reported at the
radius $r_{500c}$ enclosing overdensities with respect to the critical density
at the redshift of the output (below, we also use a higher overdensity level,
2500).  The corresponding velocity
\begin{equation}
v_{500c} \equiv \sqrt{\frac{GM(<r_{500c})}{r_{500c}}}
\end{equation} 
is the circular velocity at $r_{500c}$.  

The gas motions and mass measurement biases are expected to depend on
the dynamical state of clusters.  We therefore consider dynamically relaxed
and non-relaxed clusters separately in this work.  Following \citet{nag07a},
our relaxed subsample is identified based on the overall structural morphology
of their \emph{Chandra} images, mimicking the procedure used by observers.
Specifically, we visually examine mock $100$~ksec images and identify
``relaxed'' clusters as those with regular X-ray morphology and no secondary
maximal and minimal deviations from elliptical symmetry.  By contrast,
``unrelaxed'' clusters are those with secondary maxima, filamentary X-ray
structures, or significant isophotal centroid shifts (see Figure~1 of
\citet{nag07a} for the typical examples of systems classified as relaxed or
unrelaxed).

In order to assess the effects of gas cooling and star formation, we
also repeated each cluster simulation with only the standard
gasdynamics for the baryonic component without radiative cooling and
star formation.  We will use labels 'cooling+SF' (CSF) and
'non-radiative' (NR) to refer to these two sets of runs, respectively.
Our main analyses are based on the CSF runs.  We present results of
the NR runs only in the Section~\ref{sec:adcsf} and
Fig.~\ref{fig:mfracpro_ad} and~\ref{fig:dcc_m500_ad}.

\begin{figure}
\begin{center}
\epsscale{1.2} \plotone{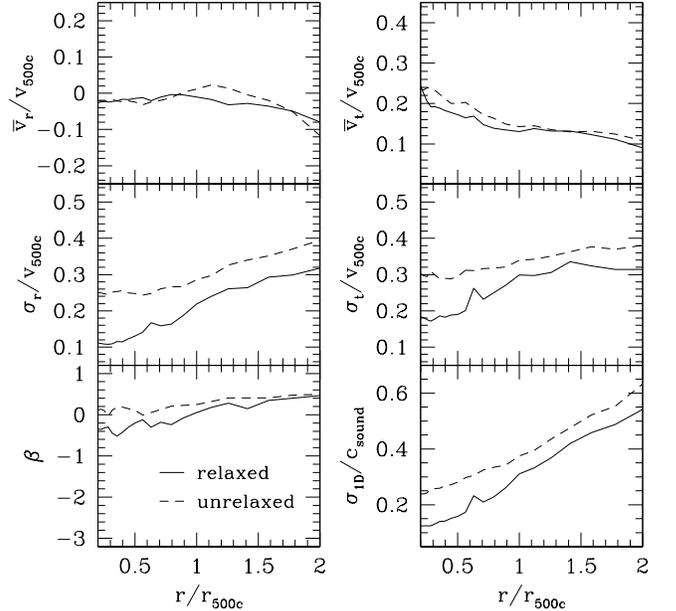}
\caption{
  Profiles of the ICM mean radial and tangential velocity, radial
  velocity and tangential velocity dispersion, the
  anisotropy parameter $\beta$, and ratio of the 1D velocity dispersion to the
  sound speed of gas (Mach number).  The solid lines represents relaxed clusters
  and dashed lines represents unrelaxed clusters.  \label{fig:icm_profiles}}
\end{center}
\end{figure}

\section{Results}
\subsection{Velocity Profiles}
\label{sec:vprof}

Following \citet{faltenbacher_etal05} we analyze the velocity profiles
of ICM for the sample of 16 simulated clusters at $z=0$.  Our
definition of the cluster center differs from that of
\citet{faltenbacher_etal05}: we use the location of the particle with
the largest local matter density.  The mean velocities in spherical
coordinates ($\bar{v}_r$,$\bar{v}_\theta$,$\bar{v}_\phi$) on each
radial shell are computed as mass-averages of gas velocities randomly
sampled on the shell where we take the rest frame of the shell to be
the peculiar velocity of the cluster (defined as the mass-weighted
average of DM particle velocities enclosed within
$r_{500c}$). Changing the rest frame to that of the gas shell does not
introduce significant changes to our analyses.  We also rotate the gas
shell such that the z-axis of the shell is aligned with the gas
angular momentum vector of the shell.  We quantify the 
random gas motion using the trace of the velocity dispersion tensor
\citep{binney_tremaine08}:
\begin{equation}
\sigma_{ij}^2 = \overline{v_iv_j}-\bar{v}_i\,\bar{v}_j
\end{equation}
where $i$,$j$ represent the spatial coordinate indices and the
overline denotes mass-average over the gas shell.  Throughout this
paper, we take random gas motions to be synonymous with random gas
motions.

While we find no significant variation in the velocity profiles with
cluster mass, the gas velocity structure between the relaxed sample
and the unrelaxed sample are different.  In the {\it top} panels of
Figure~\ref{fig:icm_profiles} we show the profiles of mean gas radial
velocity ($\bar{v}_r$) and the mean gas rotational (tangential)
velocity $\bar{v}_{t}$.  For both relaxed and unrelaxed samples, the
mean radial velocity is negative and small ($|\bar{v}_r|<0.05
v_{500c}$) up to $r_{500c}$.  Beyond this radius, the radial
velocities become increasingly more negative with radius, indicating
the radial infall of gas into the cluster potential. The mean
rotational velocity is small ($\sim 0.2 v_{500c}$ at $r=0.5 r_{500c}$)
but a bit larger than the mean radial velocity, and decreases with
radius for both relaxed and unrelaxed samples.  The {\it middle}
panels of Figure~\ref{fig:icm_profiles} show the random motions of the
radial and rotational components. Random motions in the unrelaxed
sample is in general higher overall than the relaxed sample in both
velocity components. Radial random motion increases with radius for
both samples, and the relaxed sample rises more steeply than the
unrelaxed one. Rotational random motions also increases with radius
but less rapidly that they are approximately constant over the radial
range shown. Also notice that the random motions in both radial and
tangential components are about 10 times higher than their mean
counterparts.
  
For both samples the internal gas flow remains subsonic throughout the
clusters, but the Mach number $\mathcal{M}$ \footnote[\dag]{The Mach
number $\mathcal{M}$ of the flow is given by the ratio of the
one-dimensional gas velocity dispersion
($\sigma_{\mathrm{1D}}=\sqrt{(\sigma_r^{2}+\sigma_t^{2})/3}$) to the
sound speed of gas ($c_{\mathrm{sound}}=\sqrt{\gamma
P_{\mathrm{ther}}/\rho_{\mathrm{gas}}}$).}  of the flow increases with
radius. Shown in the {\it bottom-right} panel of
Fig.~\ref{fig:icm_profiles}, for unrelaxed clusters the Mach number
increases from $\mathcal{M} \sim 0.3$ at $r=0.4r_{500c}$ to $0.5$ at
$r_{500c}$ and $\sim 0.65$ at 2$r_{500c}$. The Mach number for relaxed
clusters is in general low compared to the unrelaxed clusters, and it
increases from $\mathcal{M} \sim 0.15$ at $r=0.3r_{500c}$ to $\sim
0.4$ and $0.55$ at $r=r_{500c}$ and 2$r_{500c}$, respectively.

Some level of anisotropy is expected in reality. It is therefore
important to examine the radial component and its tangential component
of the 3D velocity dispersion.  The anisotropy parameter $\beta$
\citep{binney_tremaine08}
\begin{equation}
\beta(r) = 1 - \frac{\sigma^2_t(r)}{2\sigma^2_{r}(r)}.
\label{eq:ani} 
\end{equation}  
provides a useful measure of the relative importance of the radial and
tangential velocity components. There is a significant difference in
gas velocity anisotropy between the relaxed and unrelaxed sample.  In
relaxed clusters, near the core of the cluster the gas velocity is
slightly more tangential, with $\beta \approx -0.8 $ at $r\approx
0.2r_{500c}$. The gas velocity becomes increasingly radial as one goes
further away from the cluster center, with $\beta$ rising to nearly
zero at $r_{500c}$.  Unrelaxed clusters, on the other hand, exhibit
isotropic gas velocity at small radii and slowly becoming more mildly
radial as radius increases.

\begin{figure}
\begin{center}
\epsscale{1.0}
\plotone{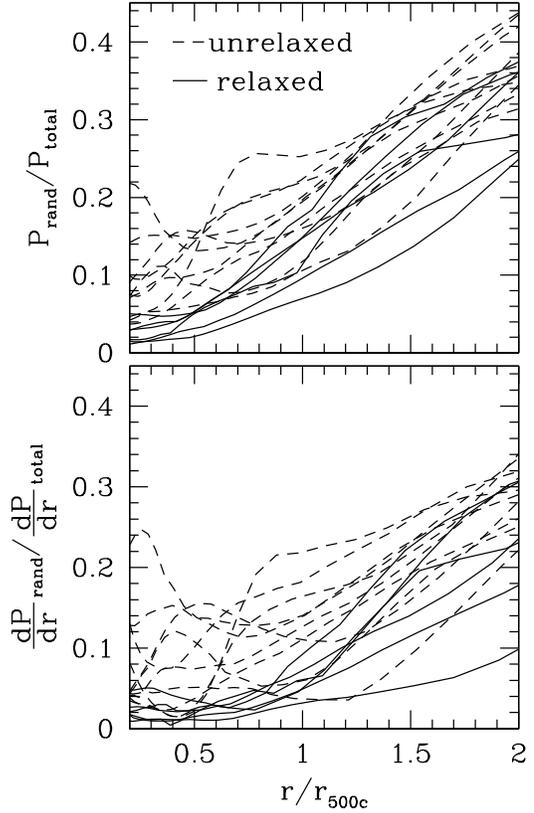}
\caption{
    Top: Ratio of pressure from random gas motions to total pressure as a function of radius.
    Relaxed clusters are represented by solid lines while unrelaxed clusters are
    represented by dashed lines.  Bottom: Ratio of pressure from random gas motions gradient to
    total pressure gradient for the 16 clusters.  Relaxed clusters are
    represented by solid lines while unrelaxed clusters are represented by
    dashed lines. 
  \label{fig:pprofiles}}
\end{center}
\end{figure}

\subsection{Pressure From Random Gas Motions in Clusters}
\label{sec:pturb}

To gauge the relative importance of pressure support from random gas motions in clusters, we
measure the ratio of the isotropic pressure from random gas motions $P_{\rm rand}$, given by  
\begin{equation}
P_{\mathrm{rand}} = \frac{1}{3}\rho_{\mathrm{gas}}\left(\sigma_r^{2}+\sigma_t^{2}\right), 
\label{eq:pturb}
\end{equation}
to the total pressure, $P_{\rm tot}\equiv P_{\rm th}+P_{\rm rand}$, as
a function of the cluster-centric radius for all 16 clusters, where
$P_{\rm th}$ is the thermal pressure of gas. Note that random gas
pressure and its gradient is sensitive to small-scale clumps and any
pressure inhomogeneity, and these sources could potentially bias the
measurements of the pressure gradient and hence the hydrostatic mass
estimate.  To minimize such bias, we remove subhaloes with mass
greater than 10$^{12}$ $h^{-1} {\rm M_{\odot}}$ and the mass within
their tidal radius from the calculation.  Lowering this mass threshold
further to 10$^{9}$ $h^{-1} {\rm M_{\odot}}$ (which excludes all
subhaloes) or increasing the radius around subhaloes within which
material is removed produces no differences in our results.  In
addition, we smooth the pressure and pressure gradient profiles using
a Savitzky-Golay (SG) filter with a smoothing scale $l = 0.95r$ where
$r$ is the distance from the cluster center and second order
polynomial interpolation \citep{numrec}. The choice of $l$ and the
polynomial order was a result of experimentation and allows to reduce
the noise in the profiles without oversmoothing them.

\begin{figure}
\begin{center}
\epsscale{1.0} 
\plotone{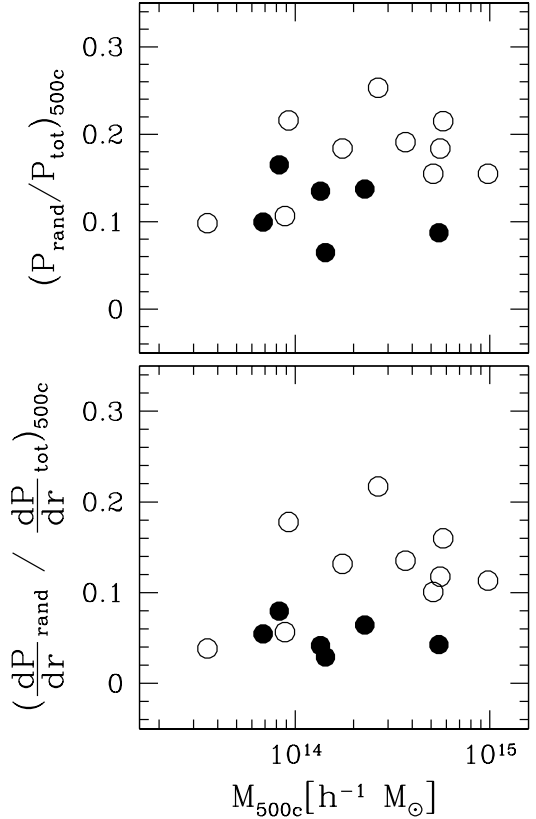}
\caption{ Top: Plot of pressure from random gas motions fraction at $r_{500c}$
  versus cluster mass within $r_{500c}$.  Bottom: Plot of fraction of
  pressure from random gas motions gradient at $r_{500c}$ versus cluster mass within
  $r_{500c}$.  Relaxed clusters are represented by filled circles
  while unrelaxed clusters are represented by empty circles.
\label{fig:pfrac_m500}}
\end{center}
\end{figure}

In the upper panel of Figure~\ref{fig:pprofiles}, we plot the fraction of the
pressure from random gas motions to the total pressure as a function of radius for
each individual cluster.  For both relaxed and unrelaxed samples,
pressure from random gas motions fraction increases with radius for $r>0.5r_{500c}$.
For relaxed clusters, the pressure from random gas motions fraction spans a range
from $6\%$ to $15\%$ at $r_{500c}$. Unrelaxed systems, on the other
hand, exhibit a greater pressure from random gas motions fraction on average, and
with larger scatter ranging from $9\%$ to $24\%$ at $r_{500c}$. (See
also Figure~\ref{fig:pfrac_m500}.)
 
The stability of the ICM in the cluster potential is provided by the
gradient of the gas pressure, and it is this gradient that is directly
relevant to the hydrostatic mass estimates. Assuming spherical
symmetry, in the bottom panel of Figure~\ref{fig:pprofiles} we plot the ratio
of the pressure from random gas motions gradient ($dP_{\rm rand}/dr$) to that of the
total pressure ($dP_{\rm tot}/dr$) for all clusters.  For relaxed
clusters, the pressure gradient fraction is relatively constant within
$0.5r_{500c}$ and increases slowly with radius. At $r_{500c}$,
random gas motions contributes to about $4\%-10\%$ of the total pressure
gradient. For unrelaxed clusters, the pressure gradient is larger,
with a range of $3\%$ to $18\%$. Differences among relaxed and
unrelaxed systems are more pronounced in the gradient than the
pressure itself.

As we showed in \S~\ref{sec:vprof}, the random gas motions are
generally anisotropic. The isotropic pressure from random gas motions expressed in terms of 
radial velocity dispersion and the anisotropy parameter is
\begin{equation}
P_{\mathrm{rand}} = \rho_{\mathrm{gas}}\sigma_r^2\left(1-\frac{2}{3}\beta\right). 
\end{equation}
This means a purely radial pressure from random gas motions is underestimated by
$2\beta/3$ if we assume isotropy.  At the scale of our interest $r
\sim r_{500c}$ where the anisotropy parameter is nearly zero for both
relaxed and unrelaxed samples, the pressure fraction profiles remain
very similar to Figure~\ref{fig:pprofiles} when we use radial random
motions to calculate pressure from random gas motions.  
   
\begin{figure}
\begin{center}
\epsscale{1.0}
\plotone{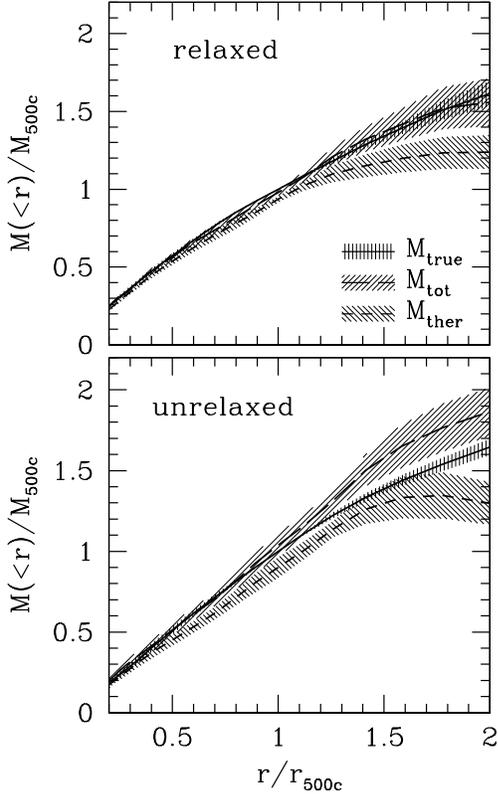}
\caption{
  Averaged mass profiles $M(<r)$ normalized by $M_{500c}$, for relaxed
  clusters (top) and unrelaxed clusters (bottom).  The solid line shows the
  actual mass profile from simulation, the long dashed line shows the mass
  profile from hydrostatic equilibrium including random gas and thermal
  pressure, and the short dashed line shows the mass profile from hydrostatic
  equilibrium including thermal pressure only.  
  Hashed region shows the 1-$\sigma$ error of the mean. 
\label{fig:mpro}}
\end{center}
\end{figure}

\begin{figure}
\begin{center}
\epsscale{1.0}
\plotone{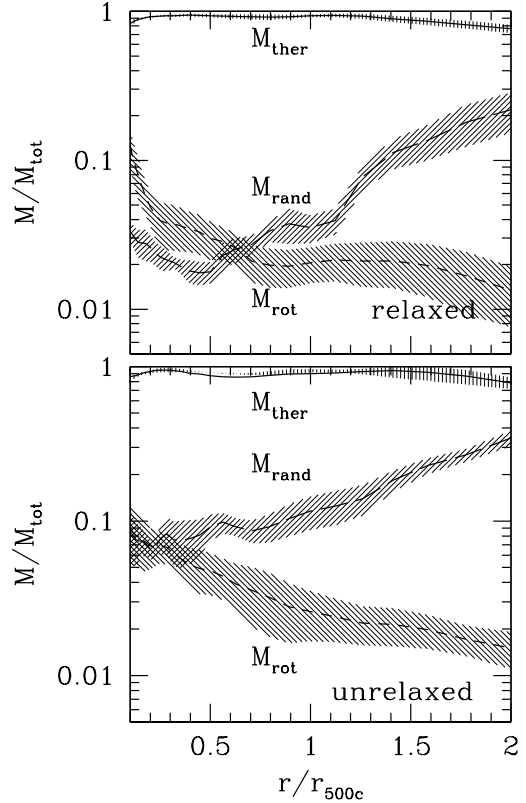}
\caption{Averaged mass fraction of various pressure contributions for relaxed
  clusters (top) and unrelaxed clusters (bottom).  The solid line
  shows the contribution from thermal pressure, the long dashed line
  shows the contribution from pressure from random gas motions, and the short
  dashed line shows the contribution from rotational support. 
  Hashed region shows the 1-$\sigma$ error of the mean. 
\label{fig:mfracpro}}
\end{center}
\end{figure}

The relative contributions of the pressure from random gas motions and
its gradient at $r_{500c}$ as a function of cluster mass $M_{500c}$ are shown in 
Figure~\ref{fig:pfrac_m500}, which shows that there is no statistically significant trend with mass
for both. The figure illustrates again the fact that in 
relaxed clusters random gas motions has smaller contribution to the total pressure gradient
compared to the unrelaxed systems.
If gas motions is indeed stirred up by dynamical interactions between haloes, 
then random gas motions should have subsided for relaxed systems via 
dissipative relaxation processes.

\subsection{Effect of random gas motions on the hydrostatic mass estimates}  
\label{sec:mbias}

In this section we assess the effects of random gas motions on the
hydrostatic mass estimate of galaxy clusters. 
We treat the random gas motions using approach similar to 
treatment of dynamical collisionless system using the 
the Jeans Equation \citep{binney_tremaine08}. 
The mass of the cluster within a radius $r$ can be split into separate
components corresponding to different terms of the Jeans equation: 
\begin{equation}
M_{\rm tot}(<r) = M_{\rm th}+M_{\rm rand}+M_{\rm rot}+M_{\rm stream}+M_{\rm cross}.
\end{equation}
The support from thermal pressure of the ICM (the HSE mass) is
\begin{equation}
M_{\rm th}(<r) = \frac{-r^2}{G \rho_{\rm gas}}\frac{dP_{\rm th}}{dr};
\end{equation}
the support from random gas motions is
\begin{equation}
M_{\rm rand}(<r) = \frac{-r^2}{G \rho_{\rm gas}}\left(\frac{\partial(\rho_{\rm gas}\sigma_r^2)}{\partial r}\right)-\frac{r}{G}\left(2\sigma_r^2-\sigma_t^2\right);
\end{equation}
the support from non-random gas rotation is
\begin{equation}
M_{\rm rot}(<r) = \frac{r\bar{v}_{t}^2}{G};
\end{equation}
the ``streaming'' term is given by
\begin{equation}
M_{\rm stream}(<r) =-\frac{r}{G}\left(r\bar{v}_r\frac{\partial \bar{v}_r}{\partial r}
+\bar{v}_\theta\frac{\partial \bar{v}_r}{\partial \theta}
+\frac{\bar{v}_\phi}{\sin\theta}\frac{\partial \bar{v}_r}{\partial \phi}\right);
\end{equation}
and the ``cross'' term
\begin{eqnarray}
M_{\rm cross}(<r) &= \frac{-r^2}{G \rho_{\rm gas}} \left(\frac{1}{r}\frac{\partial(\rho_{\rm gas}\sigma_{r\theta}^2)}{\partial \theta}+\frac{1}{r\sin\theta}\frac{\partial(\rho_{\rm gas}\sigma_{r\phi}^2)}{\partial \phi}\right) \nonumber \\
&- \frac{r}{G}\left(\frac{\cos\theta}{\sin\theta}\sigma_{r\theta}^2\right). 
\end{eqnarray}
The last two mass terms are much smaller than the other terms and can
in principle be neglected, but we include them in the actual
computation for completeness. Note that spherical symmetry for
the gravitational potential and steady state are assumed in
deriving the expression for mass from the Jeans equation, and all the
physical quantities at a given radius are averages over a radial shell.

Figure~\ref{fig:mpro} compares the true and estimated averaged
mass profiles normalized by $M_{500c}$ for relaxed and
unrelaxed clusters. The comparison shows that the HSE mass
profile $M_{\rm ther}$ ({\it short-dashed line}) underestimates 
the true mass profile measured directly in simulations ({\it solid line}).  
The mass profile from the full mass expression with both thermal and gas motion support ({\it
long-dashed line}) recovers the true cluster mass profiles quite
well for relaxed clusters. For unrelaxed clusters, the non-thermal terms 
recover the true mass well within $r_{500c}$, but overestimates the true mass 
beyond $r_{500c}$, because the assumptions of spherical symmetry does not hold 
true at large radii in most unrelaxed clusters.  

Figure~\ref{fig:mfracpro} shows the relative contribution of mass terms $M_{\rm th}$, 
$M_{\rm rand}$ and $M_{\rm rot}$. For both relaxed and unrelaxed samples, 
the thermal component dominates the pressure support ($\sim 80-90\%$ at $r_{500c}$). 
The support due to gas motions in the unrelaxed sample is higher than that of the relaxed one. 
For both samples, the rotational term is greater than the random gas term
in the inner region ($r<0.6r_{500c}$ for relaxed sample and $r<0.2r_{500c}$ for unrelaxed sample),
and decreases with radius.
The random gas term on the other hand increases
with radius, and accounts for most of the non-thermal gas support beyond $r_{500c}$. 
This generally agrees with the findings of \citet{fang_etal09} which use the same set of clusters,
despite our assumption of spherical symmetry which they do not adopt. 

\begin{figure}
\begin{center}
\epsscale{1.0} \plotone{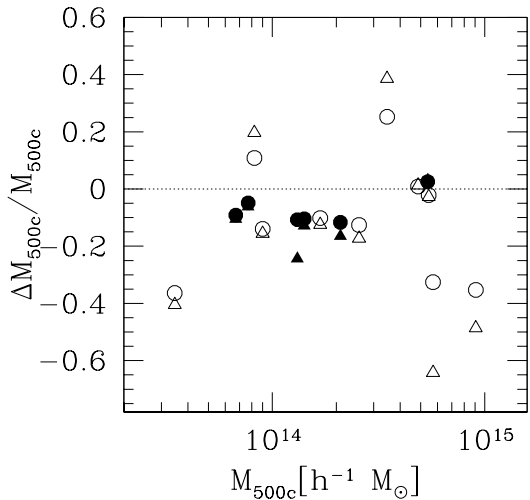}
\caption{Fractional differences between the true mass and the HSE
  estimated mass, $\Delta M/M \equiv (M_{\rm est}-M_{\rm true})/M_{\rm true}$, 
  as a function of cluster mass $M_{500c}$.  The circles and triangles show
  the hydrostatic mass evaluated at the true and estimated $r_{500c}$,
  respectively.  The solid and open symbols indicate relaxed and
  unrelaxed clusters. 
 \label{fig:dmm_m500}}
\end{center}
\end{figure}

The biases in the hydrostatic mass estimate at $r_{500c}$ as a
function of cluster mass are shown in Figure~\ref{fig:dmm_m500}.  The bias
($\Delta M/M$) is defined as a fractional difference between the
estimated mass ($M_{\rm est}$) and the true mass ($M_{\rm true}$)
measured directly in the simulations.  Following \citet{nag07a}, the
biases are computed in two ways for each cluster.  First, both $M_{\rm
est}$ and $M_{\rm true}$ are measured in the same physical region
enclosed within ``true'' radii ($r_{\rm true}$) measured directly in
simulations. The bias computed this way is indicated with circles.  In
practice, additional biases in the estimated masses ($M_{\rm est}$)
could arise from a bias in the estimation of a cluster virial radius.
Therefore, we also compute the bias in the estimated hydrostatic mass,
$M_{\rm est}(<r_{\rm est})$, enclosed within the estimated virial
radius, $r_{\rm est}$, indicated with triangles.  Both bias values
are listed in Table~\ref{tab:mbias} at $z=0$ and $z=0.6$. 
For relaxed clusters, we find that the hydrostatic mass at $r_{500c}$ is biased
low by $7\%$ and $11\%$ at $z=0$ and $0.6$, respectively.  The bias is
smaller if the hydrostatic mass is measured in the inner region of
clusters (e.g., $r_{2500c}$).  Both bias and scatter are larger for
unrelaxed systems. We find no apparent trend of the bias with cluster
mass.

Note that the biases in hydrostatic mass reported in this work are
smaller than the values reported in \citet{nag07a} based on analyses
of mock Chandra observations. Although these values are consistent
within 1-sigma, we find systematic deviations in the mean.  For the
relaxed clusters at $z=0$, the bias at $r=r_{500c}$ is 7\% which is
smaller than 13\% reported in \citet{nag07a}.
Figure~\ref{fig:dcc_m500} illustrates that the distribution of the HSE
mass in \citet{nag07a} is skewed towards more negative bias relative
to the results of the present work. The offset is smaller in the
inner regions ($\sim 1\%$ at $r=r_{2500c}$). The main difference is
that biases reported in \citet{nag07a} were determined in the
analyses of the mock {\sl Chandra} data using analysis procedures
similar to those used in observations, while the present analysis
uses the 3D profiles of gas density and temperature measured directly
from raw simulation data. We find that the difference in the inferred
bias is
primarily due to differences in the temperature profiles slope derived 
from the mock Chandra analyses in \citet{nag07a} and the 3D mass-weighted
temperature profiles directly measured from simulations.

\begin{table}[tp]
\begin{center}
\caption{Bias in the hydrostatic mass estimate of clusters}\label{tab:mbias} 
\begin{tabular}{ c c c c c } 

\hline
\hline
 & & Sample & \multicolumn{2}{c}{Bias~$\pm$~Error (1$\sigma$)
\footnote[\ddag]{The scatter can be obtained from multiplying the error by $\sqrt{N-1}$
where $N$ is the number of clusters.}} \\
\cline{4-5}\\[-1.7ex]
redshift & radial range & (number of clusters) & $\Delta M(<\!r_{\rm true})/M$ & $\Delta M(<\!r_{\rm est})/M$ \\
\hline
    &                       & all (16) & -0.077$\pm$0.022   & -0.113$\pm$0.033 \\
$z=0$ & $<r_{2500c}$ & relaxed (6) & -0.061$\pm$0.023   & -0.080$\pm$0.031 \\
    &                       & unrelaxed (10) & -0.086$\pm$0.032   & -0.133$\pm$0.049 \\ 
\hline   
    &                       & all (16) & -0.094$\pm$0.040   & -0.131$\pm$0.062 \\
$z=0$ & $<r_{500c}$  & relaxed (6) & -0.074$\pm$0.022   & -0.112$\pm$0.038 \\
    &                       & unrelaxed (10) & -0.106$\pm$0.064   & -0.142$\pm$0.098 \\
\hline   
\hline                                                                         
      &                      & all (16) & -0.066$\pm$0.023 & -0.074$\pm$0.041 \\
$z=0.6$ & $<r_{2500c}$ & relaxed (6) & -0.042$\pm$0.027 & -0.044$\pm$0.052 \\
      &                      & unrelaxed (10) & -0.081$\pm$0.034 & -0.093$\pm$0.059 \\
\hline  
      &                      & all (16) & -0.116$\pm$0.029  & -0.200$\pm$0.045 \\
$z=0.6$ & $<r_{500c}$ & relaxed (6) & -0.111$\pm$0.027  & -0.154$\pm$0.039 \\
      &                      & unrelaxed (10) & -0.119$\pm$0.044  & -0.227$\pm$0.068 \\
\hline
\hline
\end{tabular}
\end{center}
\end{table}

\begin{figure}
\begin{center}
\epsscale{1.0}
\plotone{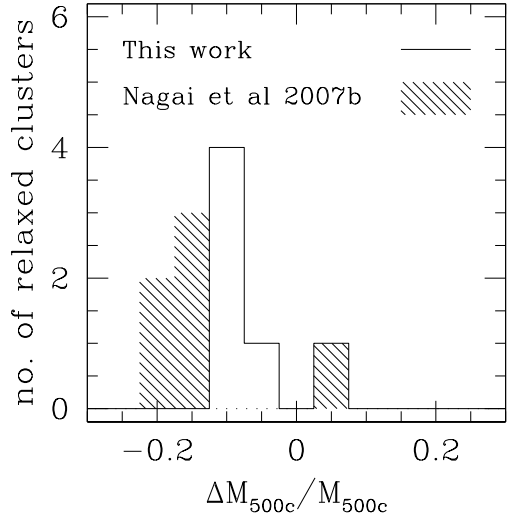}
\caption{Distributions of the HSE mass bias, $\Delta M/M \equiv
(M_{\rm est}-M_{\rm true})/M_{\rm true}$, for a sample of 6 relaxed
clusters at z=0.  The solid line shows the distribution where the
HSE mass is calculated directly from the gas thermal pressure and
density profiles. The shaded area shows the distribution where the
HSE mass is derived from mock Chandra analyses of simulated clusters
in \citet{nag07a}.
\label{fig:dmm_dist}}
\end{center}
\end{figure}

\subsection{Effect on Concentration Parameters of Galaxy Clusters}
\label{sec:conc}
\begin{figure}
\begin{center}
\epsscale{1.0} \plotone{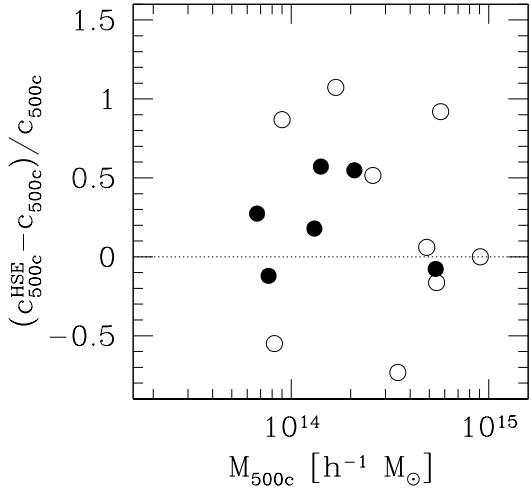}
\caption{Fractional difference of concentration $c_{500}$ derived from
the fit to the true total density profiles of clusters and to the
HSE-derived density profiles in the simulations with cooling for relaxed
(filled circles) and unrelaxed (open circles) clusters.
\label{fig:dcc_m500}}
\end{center}
\end{figure}

As seen from Figure~\ref{fig:mpro}, the bias in the HSE mass increases
with radius, indicating that the contribution of the random gas
pressure increases toward cluster outskirts.  This results in the
total density profile that is steeper than the true density profile
and hence an overestimate of concentration parameter $c_{500} \equiv
r_{500c}/r_s$ if the derived density profiles of clusters are fitted
with the NFW profile \citep{nfw96}.  Figure~\ref{fig:dcc_m500} shows
the bias in the concentration parameters for our simulated clusters.
The concentrations were estimated by fitting the NFW profile to the
derived total density profile over radial range\footnote{The lower
limit on the radius is chosen not to include the region of the profile
significantly affected by the dark matter contraction in response to
baryon condensation \citep{gnedin_etal04}, while the outer radius
corresponds to the largest radius to which the density profile is
determined reliably from current X-ray data.} $0.1\leq r/r_{500c} \leq
1.0$. We find that the estimated concentration parameter based on the
HSE assumption is biased high on average by about $(23\pm 10)\%$ for
relaxed systems and $(47\pm 21)\%$ for unrelaxed systems. The scatter is $(23\pm 5)\%$
and $(41\pm 12)\%$ for relaxed and unrelaxed systems respectively.

Given that the bias in the total mass profile we find from the
idealized analysis of the 3D cluster profiles is lower than the bias
derived from the analyses of mock {\sl Chandra} data from the same
simulation (see discussion in the previous section), we can expect
that the bias in concentrations may also be somewhat larger for the
mock data analysis. Therefore we have fitted NFW concentrations to the
total density profiles in the same radial range as above derived from
mock X-ray data for relaxed clusters used in \citet{nag07a}.  The
concentration parameters in this case tend to be overestimated by
about a factor of two.  One particular problem we encountered with
fits to some of the systems was that errors in estimates of
concentrations can become catastrophic if the NFW scale radius of the
fit was smaller than the minimum radius of the radial range used in
the fit (which corresponds to concentration values of $>10$ for our
choice of the radial range).

The upward bias of the concentrations estimated from the HSE-derived
density profiles due to gas motions can help explain the surprisingly
high values of concentrations measured in some of the recent X-ray
analyses \citep{maughan_etal07,buote_etal07}.  In particular,
\citet{buote_etal07} found that for a given mass concentrations
derived for groups and clusters from HSE X-ray analysis are somewhat
higher than expected in the standard $\Lambda$CDM model with
$\sigma_8\approx 0.7-0.8$.  The bias due to gas motions can explain
part of this difference.\footnote{Simulations also predict that
condensation of baryons and their conversion into stars should also
increase concentrations compared to the purely dissipationless
simulations \citep{rudd_etal08}, which are used to make predictions
for the concentration--mass relation in a given cosmology.}
As noted by \citet{fang_etal09}, the actual values of concentration bias
may be affected by the overcooling problem of cluster simulations. However, 
the existence and sign of the bias are generic. In particular, as we show
in the next section, qualitatively similar bias is obtained from the
simulations of our cluster sample in non-radiative regime.

\begin{figure}
\begin{center}
\epsscale{1.0}
\plotone{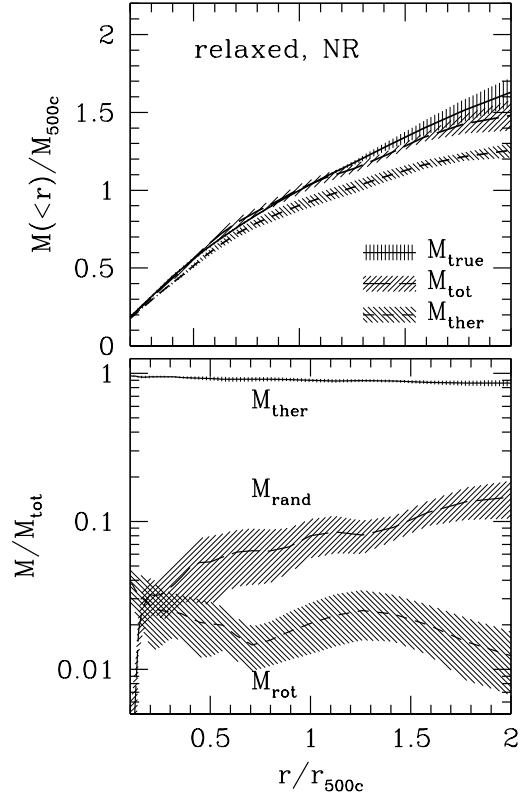}
\caption{Top panel: the mass profiles (top) for relaxed clusters
  in the {\it non-radiative} re-simulations of clusters used in our analysis; bottom panel: 
  contribution of thermal pressure, rotational and random gas motions to the total 
  mass profile for the clusters in the upper panel. 
  The notation is the same as Figure~\ref{fig:mpro} 
  and Figure~\ref{fig:mfracpro}. \label{fig:mfracpro_ad}}
\end{center}
\end{figure}

\begin{figure}
\begin{center}
\epsscale{1.1} \plotone{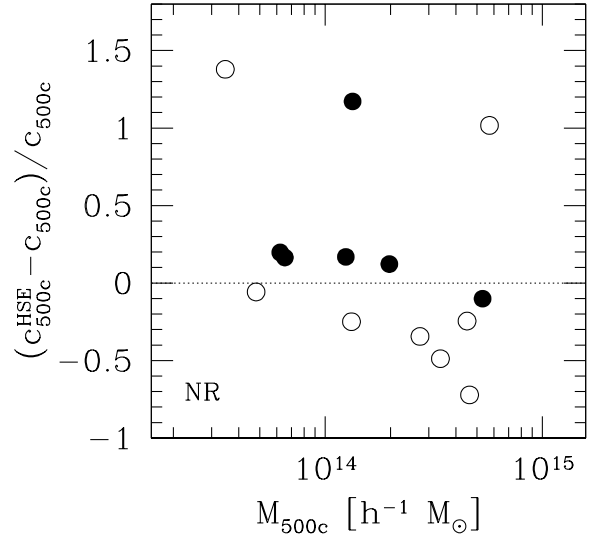}
\caption{Fractional difference of concentrations $c_{500c}$ derived
from the true total density profile and the HSE-derived density
profiles for the non-radiative re-simulations of our clusters.  The
filled circles represent relaxed clusters and the open circles
represent unrelaxed clusters.  As in the runs with cooling, the
concentrations of the relaxed clusters tend to be biased high due to
the bias in total mass, which increases with cluster-centric radius.
\label{fig:dcc_m500_ad}}
\end{center}
\end{figure}

\section{Effects of cooling on the mass and concentration biases}
\label{sec:adcsf}

Given that the cosmological simulations suffer from the ``overcooling
problem'' -- a too high a fraction of cooled, condensed gas produced in
simulations compared to observations \citep[e.g., see Fig. 2 in][]{kravtsov_etal09} -- 
it is reasonable to ask whether the results we presented in the previous
sections are affected by it. \citet{fang_etal09} have independently 
analyzed the same set of simulations we use here and argued that much of
the non-thermal pressure support in the simulations with cooling is due
to rotation of gas in the inner regions ($r\lesssim 0.5r_{\rm vir}$), which
should not be trusted because rotation is overestimated due to the overestimated
amount of cooling gas. 

Figure~\ref{fig:mfracpro_ad} shows the mass profiles for relaxed
clusters in non-radiative re-simulations of the 16 clusters used in
our analysis and contributions to the total mass from thermal pressure
and rotation and random gas gas motions. The figure shows that while
the relative contribution of rotational velocity in these simulations
is lower in non-radiative simulations, the {\it total} contribution of gas
motions to the HSE reconstruction of the mass profile is similar to
that of the simulations with cooling. Given similar degree of gas
motions, albeit with a different contributions from rotation and
random components in the inner regions, the HSE-derived mass profile
is consequently also biased low for non-radiative simulations and the
mass bias at $r_{500c}$ is similar to that in the simulations with
cooling.

In addition, the figure shows that the bias in the mass estimate also
increases with cluster-centric radius in this case, which means that
we can expect bias in the HSE-derived concentrations for the total
density profile. Indeed, Figure~\ref{fig:dcc_m500_ad} shows that the
concentration bias is qualitatively similar in non-radiative
simulations to that in runs with cooling: five out of six relaxed
clusters have $c^{\rm HSE}_{500c}$ biased high on average by $\approx 10-20\%$ with
concentration catastrophically overestimated in one of the clusters
(compare this figure to Fig.~\ref{fig:dcc_m500}). 
The results presented in this section
clearly show that conclusions of this study with regard to the bias in
the HSE-derived total mass profile and NFW concentrations are general
and are not determined by the unrealistic cooling of the gas in the
dissipative simulations.

\section{Discussion and Conclusions}
\label{sec:conclusions}

We presented analysis of gas motions in the ICM using high-resolution
adaptive mesh refinement cosmological simulations of a sample of
sixteen galaxy clusters spanning the virial mass range of $5\times
10^{13}-2\times 10^{15}\,h^{-1}\,{\rm M_{\odot}}$. Our study focuses
on the effects of residual gas motions on the estimates of the total 
mass profiles of clusters from the hydrostatic equilibrium analysis. We
analyze systems that appear morphologically relaxed and unrelaxed in
mock {\sl Chandra} X-ray images separately to study the effects of dynamical
state of clusters on the resulting HSE mass profile.

In broad agreement with previous studies
\citep{evrard90,frenk_etal99,rasia_etal04,faltenbacher_etal05,rasia_etal06,nag07a,
piffaretti_valdarnini08,jeltema_etal08,iapichino_etl08}, we find
that gas motions contribute up to $\approx 5\%-15\%$ of the pressure
support in relaxed clusters, which leads to underestimate of the total virial
mass in the HSE analysis accounting only for the thermal pressure.
We have used re-simulations of one of the clusters with different maximum
levels of refinement and find that the magnitude of the 
pressure support due to gas motions
at the considered radii has converged. 

On average, the hydrostatic cluster mass estimate is biased low by
about $6\pm 2\%$ at $r=r_{2500c}$ and $8\pm 2\%$ at $r=r_{500c}$ for relaxed
systems, while the biases in unrelaxed clusters are about $9\pm 3\%$ and
$11\pm 6\%$ at these radii, respectively. We have tested that our results
are not affected by resolution by analyzing re-simulations of one of
the clusters with different spatial resolution. The magnitude of the effect is
consistent with the recent observational evidence of a similar bias
based on comparison of the HSE and weak lensing derived masses
\citep{mahdavi_etal08}.

We observe that the average Mach number of the ICM gas motions is rather small
$\mathcal{M}\sim 0.4$ for unrelaxed clusters, which implies that gas
motions are generally subsonic and thus do not dissipate via shocks.
This means that it may take substantial amount of time to thermalize
these motions fully, and that the time will depend largely on the
physical viscosity of the gas. Incomplete thermalization in unrelaxed
clusters leads to a lower measured X-ray temperature $T_X$ that might
bias the $M-T_X$ scaling relation with a higher normalization
\citep[e.g.,][]{mathiesen_evrard01}. On the other hand,
underestimating the cluster mass from the hydrostatic equilibrium
analysis would bias the $M-T_X$ relation to lower masses at a given
temperature.  This issue can potentially be resolved by comparing the
HSE-derived masses to cluster masses derived from weak lensing surveys
\citep{mahdavi_etal08,zhang_etal08,vikhlinin_etal09}. 

The bias in the total mass profile increases with increasing cluster-centric radius radius which
results in the HSE-derived total density profile that is more
concentrated that the true profile. The best fit NFW concentrations
fit to the HSE-derived profile therefore results in concentration
values generally biased high compared to the corresponding fits to the
true density profiles.  In particular, we find that the concentration
parameters, $c_{500} \equiv r_{500c}/r_s$, based on the HSE mass
profile, is biased high on average by about $\sim 24\%$.  This bias
would have to be taken into account for observational studies of the
concentration-mass ($c-M$) relation based on the HSE assumption
\citep{maughan_etal07,buote_etal07}. We show that a similar bias exist
in re-simulations of the same clusters in the non-radiative regime. Our
conclusions therefore are general and are not due to the overcooling
in the dissipative simulations, as was claimed by \citet{fang_etal09}.

We demonstrate that the mass profile within $r_{500c}$ can be
recovered well if the pressure support due to gas motions is
explicitly taken into account.  On the other hand, the recovery is not
as good beyond $r_{500c}$, with the deviation from the true mass
profiles depending on the cluster dynamical state and configuration of
mergers.  There are some uncertainties in our analysis due to
limitations of our simulations. Although we show that adding gas
motion support recovers the mass of a relaxed cluster accurately, we
have not considered other potential sources of non-thermal pressure
support, such as cosmic rays and magnetic fields, that can also bias
the hydrostatic mass measurement. These effects have to be considered
separately.  Note that the effect of gas motions generally increases
with increasing radius simply because clusters are less relaxed at
larger radii. The effects of other sources of non-thermal pressure
support can have different radial dependence and future observations
may potentially exploit such differences to differentiate among them.

In addition, real gas motions in the ICM plasma may also be different
from the motions in our simulated clusters due to a different physical
viscosity of the intracluster gas compared to simulations. Given that higher
viscosity would tend to damp the gas motions faster, the estimates of
their effects presented here provide an upper limit to the random gas
pressure of the ICM. Simulations exploring reasonable scenarios for
magneto-hydrodynamics effects, effects of AGN feedback, and cosmic
rays injection and evolution should provide some guidance as to the
radial dependence of contributions from magnetic fields and cosmic
rays, as non-thermal sources of pressure support
\citep[e.g.,][]{subramanian_etal06,pfrommer07,sijacki_etal08}.
 
Interestingly, we do not detect a strong dependence of the mass bias
with cluster mass, which means that the bias would affect the
normalization of the mass observable relations but not its
slope. However, the sample of simulated clusters is rather small.
\citet{jeltema_etal08} use a considerably larger sample of simulated
clusters and do detect a weak trend of the HSE mass bias with cluster
mass.  In addition, \citet{jeltema_etal08} show that it is important
to use quantitative measures of the dynamical state of the clusters
rather than the visual classification of the mock X-ray images, as was
done in our analysis (note, however, that we follow the standard
practice of such classification in observational X-ray analyses). In
agreement with \citet{jeltema_etal08}, our own analysis shows that
often clusters that appear morphologically relaxed along one
projection, would not be classified as relaxed along other projections
(the reverse is also true, as clusters that are intrinsically relaxed
may seem unrelaxed in projection due to projecting groups and clusters
along the line of sight).

Observationally, little is known about gas motions in clusters except
that in many systems there are indirect indications of bulk gas
motions associated with mergers and motions of the central cluster
galaxies \citep[see][for a review]{markevitch_vikhlinin07}.
\citet{schuecker_etal04} presented evidence for possible random gas
motions in the Coma cluster via fluctuations in the pressure map.  It
is not yet clear, however, whether this interpretation is unique and
whether the results are generic for other clusters.  The most direct
way of measuring gas motions would be to measure broadening of line
profiles of heavy ions in radio \citep{syunyaev_churazov84} or in
X-rays \citep{inogamov_sunyaev03}. In radio wavelengths this requires
very deep observations to detect the lines in emission or sensitive
observations against bright radio sources in clusters to detect the
lines in absorption (D. Marrone, priv. communication). Observations of
the lines in X-rays require sensitive high resolution X-ray
spectrometer, which have not yet been available, but may become
possible with the launch of the ASTRO-H satellite\footnote{{\tt
http://www.isas.jaxa.jp/e/enterp/missions/astro-h/index.shtm}}.
A combination of constraints from mass measurements via gravitational 
lensing,  possible future direct measurements of gas motions, 
and improved modeling of the ICM physics is thus the way to make
progress in our understanding of the mass measurement biases discussed
in this paper.

\acknowledgments We would like to thank Douglas Rudd for providing us
re-simulations for one of our clusters to evaluate numerical
convergence of our results.  E.T.L. and A.V.K. are supported by the
National Science Foundation (NSF) under grants No.  AST-0239759 and
AST-0507666, by NASA through grants NAG5-13274 and 07-ATFP07-0153, and
by the Kavli Institute for Cosmological Physics at the University of
Chicago. The cosmological simulations used in this study were
performed on the IBM RS/6000 SP4 system ({\tt copper}) at the National
Center for Supercomputing Applications (NCSA).

\bibliographystyle{apj}
\bibliography{ms}

\clearpage

\end{document}